\documentclass[aps,pre,twocolumn,superscriptaddress,floatfix]{revtex4}
\usepackage{graphicx}
\usepackage{times}
\usepackage[dvips,unicode,colorlinks,linkcolor=blue,citecolor=blue,urlcolor=blue]{hyperref}
\usepackage{amssymb}
\begin{document}

\title{Heat conductivity of DNA double helix}

\author{Alexander V. Savin}

\affiliation{Semenov Institute of Chemical Physics, Russian Academy
of Sciences, Moscow 119991, Russia}

\author{Mikhail A. Mazo}
\affiliation{Semenov Institute of Chemical Physics, Russian Academy
of Sciences, Moscow 119991, Russia}

\author{Irina P. Kikot}
\affiliation{Semenov Institute of Chemical Physics, Russian Academy
of Sciences, Moscow 119991, Russia}

\author{Leonid I. Manevitch}
\affiliation{Semenov Institute of Chemical Physics, Russian Academy
of Sciences, Moscow 119991, Russia}

\author{Alexey V. Onufriev}
\affiliation{Departments of Computer Science and Physics, 2160C Torgersen Hall,
             Virginia Tech, Blacksburg, VA 24061, USA}

\begin{abstract}
Thermal conductivity of isolated single molecule DNA fragments is of importance
for nanotechnology, but has not yet been measured experimentally. Theoretical
estimates based on simplified (1D) models predict anomalously
high thermal conductivity. To investigate thermal properties of single
molecule DNA we have developed a 3D coarse-grained (CG) model that retains
the realism of the full all-atom description, but is significantly more 
efficient. Within the proposed model each nucleotide is
represented by 6 particles or grains; the grains interact via effective
potentials inferred from classical molecular dynamics (MD) trajectories based
on a well-established all-atom potential function. Comparisons
of 10 ns long MD trajectories between the CG and the corresponding
all-atom model show similar
root-mean-square deviations from the canonical B-form DNA, and similar
structural fluctuations. At the same time, the CG model is 10 to 100 times
faster depending on the length of the DNA fragment in the simulation.
Analysis of dispersion curves derived from
the CG model yields longitudinal sound velocity and torsional stiffness
in close agreement with existing  experiments. 
The computational efficiency of the
CG model makes it possible to calculate thermal conductivity
of a single DNA molecule not yet available experimentally.
For a uniform (polyG-polyC) DNA, the estimated conductivity coefficient is
0.3 W/mK which is half the value of thermal conductivity for water. 
This result is in stark contrast with estimates of thermal conductivity
for simplified, effectively 1D chains  ("beads on a spring")  that  
predict anomalous (infinite) thermal conductivity.
Thus, full 3D character of DNA double-helix retained in the
proposed model appears to be essential for describing its
thermal properties at a single molecule level.

\end{abstract}

\maketitle

\section{Introduction}

Heat conductivity
of nanostructures is of great importance both from fundamental and
applied points of view. For example, superior thermal conductivity has been
observed in graphene \cite{balandin,ghosh} and carbon nanotubes \cite{pop},
which has raised an exciting prospect of using these materials
in thermal devices \cite{chang,yang1,yang2,wu,hu}.
Generally, one can not expect that bulk thermal properties of a
material will remain unchanged at the nanoscale: in
some nano materials such as silicon thermal conductivity is
about two orders of magnitude smaller than that of bulk crystals \cite{li},
with the reduction in conductivity
attributed to strong inelastic surface scattering.
Furthermore, some familiar physical laws such as Fourier's 
law of heat transfer that
work in bulk materials are no longer  valid on the
nanoscale \cite{chang2,maruyama,zhang,sav1}.

Deoxyribonucleic acid (DNA) is one of the most promising
nanowire materials due to the relative ease of
modifications combined with the self-assembly capability which make
it possible to construct a great variety of DNA-based nanostructures \cite{Endo2009,Joshi2010}.
While electrical conductivity of single DNA molecules
has been extensively studied, the corresponding thermal properties remain
largely unexplored. The first, and to the best of our knowledge the only 
published work so far that  attempted to measure thermal conductivity of
single molecule DNA -- DNA-gold composite \cite{kodama} -- 
gave an estimate of  150 W/mK for the
coefficient of thermal conductivity, which was conspicuously close to that
of pure gold. The study concluded that
molecular vibrations play a key role in thermal conduction
process in DNA molecule, but thermal conductivity of
single molecule DNA remained unknown.

At the same time, theoretical
approaches to the problem have met with their own difficulties. Numerical
modeling of heat transfer along carbon nanotubes and nanoribbons showed
that thermal conductivity increases steadily with the length
of the specimen \cite{chang2,maruyama,zhang,sav1}.
If one makes an analogy with 1D anharmonic chains
that always have infinite thermal conductivity \cite{Lepri97,Lepri03},
one might interpret these results as suggesting anomalously high
thermal conductivity for quasi one-dimensional nanosystems. Since at some
level the DNA double helix may also be considered as a quasi 1D system,
one wonders if the corresponding thermal conductivity is also anomalously
high, increasing with the length of the DNA molecule? 
It is possible that over-simplified 
"beads-on-spring" models of DNA are inappropriate in this context, 
and thermal properties 
of the real double helix do not exhibit the low 
dimensional anomaly in heat conductivity. 

The goal of this work is to investigate heat conductivity of single molecule 
DNA by direct modeling of heat transfer along the double helix
via classical molecular dynamics of the DNA. To accomplish this goal 
we will have to choose a level of detail that 
is computationally feasible but at the same time 
retains key properties of the fully atomistic picture
of the molecule. 

Classical molecular dynamics  (MD) simulations based on  fully
atomistic (all-atom) representations  \cite{amb1, charmm1, charmm2}(see Fig. \ref{fig1})
are among the most
widely used tools currently employed to study dynamics of the DNA
double helix \cite{Perez-07}. In these simulations
the dynamics of the atoms is governed by semi-empirical
potentials, or force-fields;  CHARMM27 \cite{charmm1, charmm2} or AMBER
\cite{parmbsc0} are the most common force-fields that
accurately reproduce a variety of structural and dynamical properties of small fragments
of canonical and non-canonical nucleic acids in water,
at least on time-scales of up to one microsecond \cite{Perez-08, Perez-07,
Cozmuta2007,
Beveridge:2004:Biophys-J:15326025,
Makarov:2002:Acc-Chem-Res:12069622,
Giudice:2002:Acc-Chem-Res:12069619,
Cheatham:1996:J-Mol-Biol:8676379,
Feig:1997,Feig:1998:Biopolymers:10699840,
Norberg:1995}.
Importantly, classical force-fields such as AMBER \cite{Case2005}
can reproduce high-level quantum mechanical calculations for
hydrogen bonding and base stacking interactions
\cite{Perez:2005:Chemistry:15977281,Sponer:2006:Chemistry:16425171}.
However, accuracy of these all-atom models in which every atom of
the DNA fragment and all of the surrounding solvent molecules
are represented explicitly comes at a price of substantial
computational expense that limits the range of applicability of the models.

The so-called {\it implicit solvent approach} \cite{
OnufrievAnnRep08,
baker05,
CramTuhlar1999,
Gilson1995,
Scarsi97}
reduces the computational
expense by replacing the discrete water environment with a continuum
with dielectric and "hydrophobic" properties of water. The solvent 
degrees of freedom are "integrated out" and the corresponding free energy 
term is added to the Hamiltonian of the system. However, even
in this case all-atom simulations may be computationally expensive.
For example, a single 5 ns long simulation of a 147 base pair
DNA fragment reported in Ref. \cite{Onufriev1}
took 115 hours on 128 processors. This example suggests that all-atom models
may not be suitable for the program set out in this work, in which
heat transfer along long fragments of DNA will have to be examined.
We therefore resort to yet another level of
approximation -- coarse-graining (CG), where sets of original atoms are grouped
into single "united atoms" particles or grains.

The remainder of this work is organized as follows.
We begin with an outline of the coarse-graining procedure
leading to the proposed model, followed by a description of
the potential function. Details are provided in the Appendix. 
We validate the model by comparing its dynamics
with that of the corresponding all-atom model. Small amplitude vibrations
and dispersion curves are analyzed next, leading to an
addition verification of the model by comparison of several
predicted characteristics (speed of sound, torsional rigidity) with the experiment.
Then, we describe in detail the formalism used to model the heat transfer
along a single DNA molecule. In "Conclusion" we
provide a summary of the results and a brief discussion.

\section{The coarse-grained model of double helical DNA}

Naturally, there is no unique prescription for
subdividing a macromolecule into grains. The grouping of individual atoms into
grains aims to achieve a balance between faithful representation
of the underlying dynamics and the associated computational expense which
is directly related to the number of grains retained in the CG description.
A fairly large number of coarse-grain DNA models has been
developed \cite{Bruant1999, Drukker2000, Drukker2001, Orozco:2003:Chem-Soc-Rev:14671790,
Mergell2003, Nielsen2004, Chen2005, Sales-Pardo2005, Tepper2005, Hyeon2006,
Errami2007, Knotts, Cadoni2007, Becker2007, McCullagh2008,  Hofler2008, Mazur2008,
Mazur-CG, Kovaleva, CG-2010}.
Many of these models are phenomenological -- each nucleotide is represented
by 1 to 3 grains interacting via relatively simple pair potentials designed
to reproduce either certain set of experimental properties or the
results of numerical simulations based on the
corresponding all-atom models.
However, the oversimplified description of the
nitrogen bases carries the risk of losing some key details of the base-base
interactions, particularly their stacking part, that affects intramolecular
rearrangements. The latter plays a very important role in
heat transfer along the DNA molecule \cite{Yakovchuk2006}.
To make sure the nitrogen bases are treated as accurately
as possible within the CG description, we follow a strategy in which
each base is modeled by three grains; the interaction between the bases
is modeled at the all-atom level via a computationally effective strategy
described below.

\begin{figure}[tbp]
\includegraphics[angle=0, width=1\linewidth]{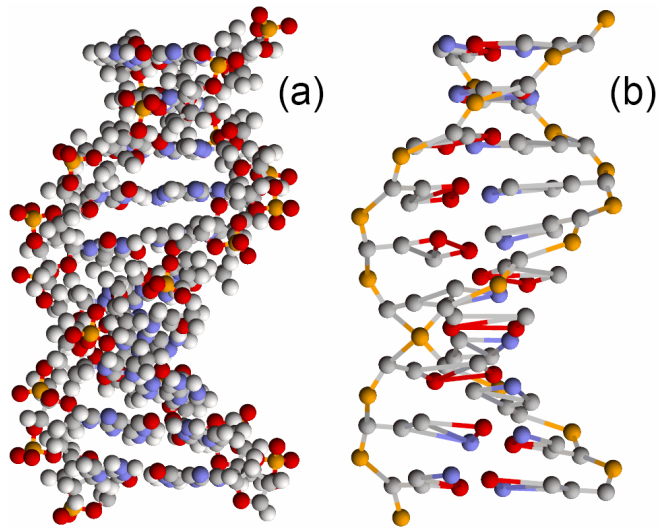}
\caption{\label{fig1}\protect
View of a DNA fragment (CGTTTAAAGC)  for (a) standard
all-atom representation
of the double helix and (b) the proposed coarse-grained model (12CG) based
on 12 united atom particles (grains) per base pair.
}
\end{figure}

\begin{figure}[tbp]
\includegraphics[angle=0, width=1\linewidth]{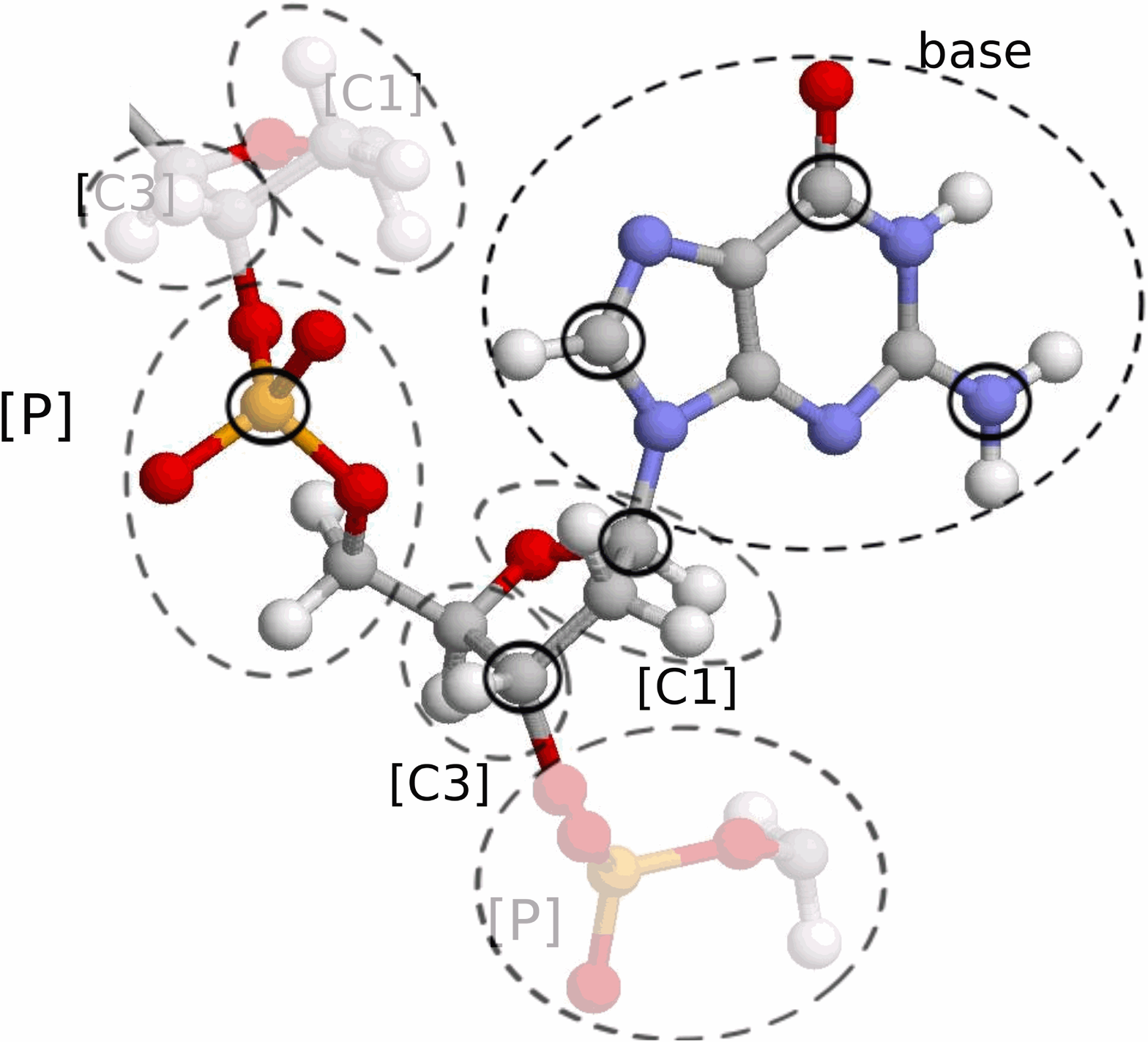}
\caption{\label{fig2}\protect
Combining original atoms into coarse grains on the DNA backbone. 
Dashed lines indicate atoms that are
included in the corresponding grain, 
solid circles mark the atoms on which the grain is centered. 
}
\end{figure}

Within the coarse-grain model each nucleotide is represented by 6 coarse-grained particles, 
or grains: 1 for the phosphate group, 2 for the sugar ring, and 3 for the 
nitrogen base.  
The mass of each coarse grain equals the net mass of the original atoms that make up that grain;  
for the 3 base grains the original mass is distributed between them as 
described in the Appendix.  
The fine-level to coarse-grain reduction employed by our model is shown in Fig. \ref{fig2}.
 Following Bruant et al. \cite{Bruant1999},
where all-atom molecular simulations were used to identify a set of relatively
rigid groups of atoms in the DNA, all of the original atoms of the phosphate
and C5$'$ groups [atoms P, O1P, O2P, O3$'$, O5$'$, C5$'$, H5$'$1, H5$'$2, see Fig. \ref{fig2}] 
are combined into a single [P] grain which is placed at the position of the original P atom.  

The sugar groups are described by two grains which are placed
on the original  C3$'$ and C1$'$ atoms;  they  will be
denoted as [C3] and [C1]. The grain  [C3] includes C3$'$, H3$'$, C4$'$ and H4$'$ original atoms, 
the grain [C1] includes original  
C1$'$, H1$'$, C2$'$, H2$'$1, H2$'$2 and O4$'$ atoms.
Thus, within our coarse-grain model the
backbone of the double helix is represented by a chain of 3 particles (grains)
[P], [C3] and [C1] (see Fig. \ref{fig2}). 

Nitrogen bases (A, T, G and C) are rather rigid, planar structures; 
spatial position and orientation of each base can
be uniquely determined from positions of any three atoms that belong
to that base. Therefore, 
bases  A, T, G, C will be described in terms of three grains. For
the A base, we identify the three grains with
the original C8, N6, C2 atoms;  for the T base, the three atoms are
C7, O4, O2; for the G they are C8, O6, N2 atoms;  and for  the C base, 
they are  C6, N4, and O2 original atoms.
Thus within the suggested model one base-pair (bp)
of the DNA double helix consists
of 12 grains -- we call the model "12CG" \ [see Fig. \ref{fig1} (b)].
For $N$ base-pair double helix, our system will consist of $12N$ particles.
Note that within our terminology the
simplest possible "beads-on-spring" \ model would be called "1CG" \
(one grain per base pair), and the all-atom representation would be "40CG"\ ,
although in this case the exact number would depend slightly on the base sequence
e.

Interactions between neighboring base pairs are obviously very important
for heat transfer along the DNA molecule. So within the framework of our
coarse-grained model the stacking of
neighboring base pairs should be taken into account  as accurately as possible.
We take advantage of the planar structure of the bases 
to bring the accuracy
of the stacking interactions close to the all-atom level, but with little
additional computational expense: from
the known grain coordinates of each coarse-grain base,
one can trivially restore
coordinates of all of the original atoms in the base with virtually
no additional computational expense. We then uses these coordinates to
calculate the stacking energy using accurate all-atom potentials,
see Appendix for details.

\section{The potential function}

To describe interactions between the grains, we employ a potential
function that contains all of the "standard" terms used in classical molecular
dynamics simulations \cite{SchlickBook02,Leach}. These terms include
internal energy contributions such as bond stretching and angle bending,
short-range van der Waals (vdW) interactions, and long-range electrostatic
interactions in the presence of water and ions. The latter are modeled
implicitly, at the continuum dielectric, linear response level. The
detailed term by term description of the potential is given in the Appendix.

The total energy of the system consists of nine terms:
\begin{equation}
H=E_k+E_v+E_b+E_a+E_t+E_{hb}+E_{st}+E_{el}+E_{vdW}.
\label{f1}
\end{equation}
The first term $E_k$ stands for kinetic energy of the system, the terms $E_v$, $E_a$, $E_t$ 
describe respectively bond, angle and torsion deformation energy of the backbone. The term $E_b$ 
stands for base deformation energy and was introduced to hold four points  -- C1$'$ and 
three points on a nitrogen base -- near one plane. Last two terms  $E_{el},E_{vdW}$ describe 
electrostatic and van der Waals interactions between grains on the backbone. 
Interaction between nitrogen bases, including interactions along the same 
chain (stacking) as well as interactions across the complementary chains (including hydrogen bonds between complementary bases), 
are described by two terms $E_{st}$ and $E_{hb}$. These 
two potentials depend on coordinates of all of the original atoms of the base. 
These coordinates are uniquely calculated 
from positions of the three grains that form each base; 
the reader is referred to Appendix for details. 
A fortran implementation of the model is freely available 
at http://people.cs.vt.edu/~onufriev/software

\section{Validation of the model}

We begin validating the proposed coarse-grain model by comparing the
resulting DNA dynamics with that produced by the corresponding well-established all-atom model. 
Later in this work we will also discuss direct comparisons
with the experiment (estimated sound velocities).

\begin{figure}[tbp]
\includegraphics[angle=0, width=1\linewidth]{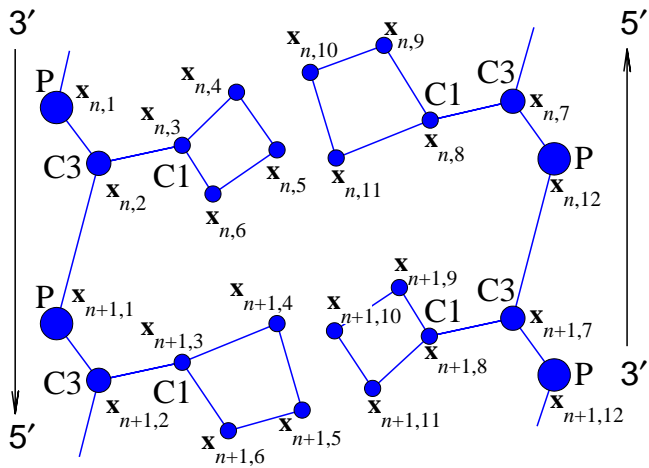}
\caption{\label{fig3}\protect
Fragment of the DNA double helix in the coarse-grained representation. 
Base-pairs $n$ and $n+1$ are shown. 
}
\end{figure}

In what follows we use following notation for convenience: 
${\bf x}_{n,j},j=1,\cdots,12$ are 
coordinates of 12 grains on the $n$-th base-pair of 
the double helix (see Fig. \ref{fig3}). 
Therefore, the configuration of $n$-th base-pair is given 
by a 36-dimensional coordinate vector 
${\bf u}_n=\{ {\bf x}_{n,j}\}_{j=1}^{12}$.   
The constant temperature dynamics of the double helix is obtained by integrating numerically the
following system of Langevin's equations:
\begin{equation}
{\bf M}_n\ddot{\bf u}_n=-\partial H/\partial{\bf u}_n -\Gamma{\bf M}_n\dot{\bf u}_n+\Xi_n,
\label{f8}
\end{equation}
where $n=1,2,...,N$, $\Gamma=1/t_r$ is the Langevin collision frequency
with  $t_r=1$ ps being the corresponding particle  relaxation time,
${\bf M}_n$ is a diagonal matrix of grain masses of $n$-th base-pair,
and $\Xi_n=\{\xi_{n,k}\}_{k=1}^{36}$ is a 36-dimensional vector of Gaussian
distributed stochastic forces describing the interaction of $n$-th base-pair
grains with the thermostat with correlation functions
$$
\langle \xi_{n,i}(t_1)\xi_{m,j}(t_2)\rangle=2M\Gamma k_BT\delta_{nm}\delta_{ij}\delta(t_2-t_1),
$$
where the mass $M=M_k$, if $i=3(k-1)+l$, $k=1,...,12$, $l=1,2,3$.

To bring the temperature of the molecule to the desired value $T=300K$,
we integrate the system (\ref{f8}) over time  $t=20t_r$
starting from the following initial conditions
\begin{equation}
\{{\bf u}_n(0)={\bf u}_n^0,~~\dot{\bf u}_n(0)={\bf 0}\}_{n=1}^N
\label{f9}
\end{equation}
that correspond to the
equilibrium state of the double helix $\{ {\bf u}_n^0\}_{n=1}^N$.
Once the system is thermalized, the temperature is maintained at $T=300K$
and the trajectory continues for 10 ns.

The first step in the validation procedure is to estimate
root-mean-square deviation (RMSd) of the end point (t=10 ns) of
the trajectory from a reference DNA structure, and compare the RMSd values
between the CG and the reference all-atom trajectory (AMBER).  Given
two structures, the RMSd can be computed as:
$$
d = \left[\frac{1}{12N} \min\limits_{{\bf S}\in SO(3),{\bf l} \in {\mathbb R}^3}
\sum\limits_{i=1}^{12N} \left( {\bf r}_i - ({\bf S}{\bf r}_{i}'+{\bf l})\right)^2\right]^{1/2},
$$
where  ${\bf r}_i, i=1,...,12N$ is the reference (e.g., initial),
and  ${\bf r}_{i}'$ is the final set of coordinates of the structure.
The expression is minimized over a translation (vector ${\bf l}$)
and a rotation around a fixed point (operator ${\bf S}$).
The details of the algorithm are described in the Ref. \cite{Horn}.
Analysis of RMS deviations from reference structures
as a function of simulation time
is commonly used as initial check of stability of the
system and quality of the underlying models \cite{Chocholousova06,Tsui-Case}.

\begin{figure}[t]
\includegraphics[angle=0, width=1\linewidth]{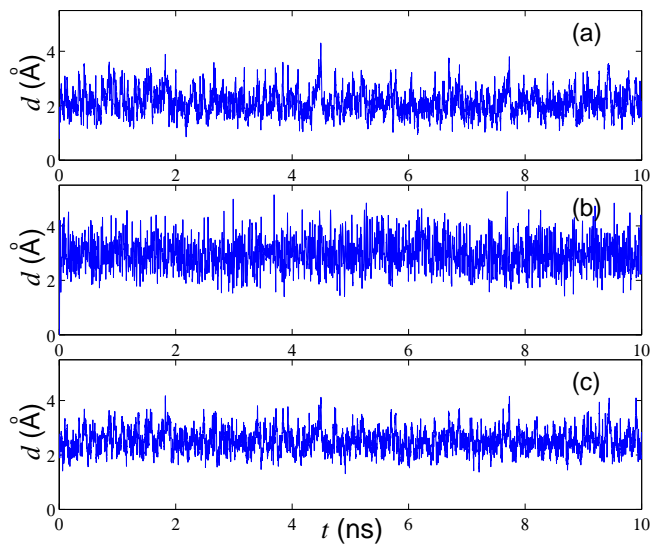}
\caption{\label{fig4}\protect
Comparison of time dependence of RMS deviation relative to
various reference structures in coarse-grained and all-atom molecular dynamics 
simulations of a 12 base-pair DNA fragment at T=300K.
(a) 12CG model simulation. RMSd is relative to the first frame. (b) All-atom
model simulation. RMSd is relative to the first frame, (c)
12CG model simulation. RMSd is relative to B-DNA X-ray
structure \cite{1BNA}. For all-atom structures the RMSd is computed only for
the subset of atoms that define grain centers in the corresponding CG model.
}
\end{figure}

As is common in the field,
the following sequence of 12 base pairs d(CGCGAATTGCGC)$_2$
(Dickerson's dodecamer) was used for this test; experimental X-ray 
structure of this B-DNA fragment is available. 
A constant temperature ($T=300K$) simulation  was performed for 10 ns.
As one can see from the Fig. \ref{fig4} the various
RMSd metrics fluctuate around their equilibrium values, which suggests that the
system remains stable in dynamics, on the time scale of the simulation.
A comparison with the corresponding all-atom simulation is shown in Fig. \ref{fig4} (b). 
This all-atom simulation uses the same 12 base-pair fragment, and
is based on the latest nucleic acid force-field (parmbsc0 \cite{parmbsc0})
from AMBER. The solvent was represented via the generalized Born implicit
solvent approximation; all other parameters such as Langevin collision
frequency, ambient salt concentration, etc. were the same as in the CG
simulation shown in Fig. \ref{fig4} (a). Comparing Figs.
\ref{fig4} (a) and (b) we can see that the all-atom RMSd is slightly
larger than that of the 12CG models. We can conclude that the 12CG model is
somewhat more rigid as compared with all-atom one.
Finally, we note that the equilibrium RMS deviation from
the experimental (X-ray) B-form
DNA is about  $2.5$\AA~, Fig. \ref{fig4} (c),
which is similar to what was observed earlier in
all-atom implicit solvent simulations \cite{Chocholousova06}.

\begin{figure}[h]
\includegraphics[width=\linewidth]{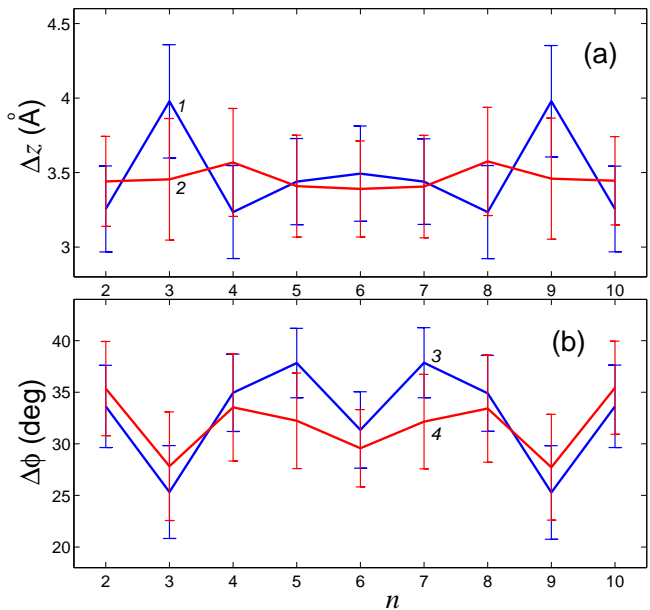}
\caption{\label{fig5}\protect
Comparison of two common helical parameters
(a) $\Delta z$ (rise) and (b) $\Delta\phi$ (twist) between the CG model
(curves 1, 3) and the corresponding all-atom model  (curves 2, 4)
($n$ -- number of base pair step). Shown are averages over the
corresponding 10ns molecular
dynamics trajectories at T=300K.
}
\end{figure}

Another common set of structural
parameters used in validation of DNA models is helical
parameters. These parameters determine the interaction between neighboring base pairs,
hence they are significant for heat transfer processes. Let's choose, for simplicity, 
two of them which are the most relevant ones for
describing the over-all structure of the double helix.
The first of these parameters is the angle $\phi$, called {\it twist},
through which each successive base pair is rotated around the helical axis
relative to its (nearest neighbor) predecessor. The second
one, {\it rise}, is the distance between such two neighboring base pairs.
Given the structure of a single nucleotide and the values of the twist and rise, one
can re-construct the whole molecule assuming that it is
a ``one-dimensional" uniform crystal.
Exact algorithm of calculating these parameters is described in \cite{x3dna}.
We used X3DNA \cite{x3dna} package and in-house software for
computing these parameters in our all-atom and CG models.
With regards to twist and rise, the validation of our 12CG
model was performed in the
same manner as previously described in the context of an all-atom
model \cite{Tsui-Case}. The results are presented in Fig. \ref{fig5},
where the averages of the 10ns simulation trajectories and the
standard deviations (indicated by error bars)
for each base pair step are
shown. One can see that the twist and rise
values for 12CG model are rather close
to those of the all-atom model.
A small difference is comparable with that seen between DNA simulations in
explicit vs. implicit solvent \cite{Tsui-Case}.

\section{The dispersion curves and small-amplitude oscillations}
The proposed 12CG model enables one to compute dynamical evolution
of a DNA molecule with any base sequence.
However, for homogeneous molecules, that is if all base pairs are identical, the
molecule can be considered as quasi-one-dimensional crystal with the
elementary cell being
one nucleotide pair of the double helix. This is a very useful simplification
that will be employed here; it is also a very reasonable one as long as the focus 
is on the over-all physics of the structure, not on sequence dependent effects. 
The main advantage of the homogeneity assumption is that
linear oscillations can be analyzed by standard
techniques of solid state physics. To be specific, let's consider a
poly-G double helical chain, assumed to extend along the $z$-axis.
In the ground state of the double helix, each successive
nucleotide pair is obtained from its predecessor by translation
along the z-axis by step $\Delta z$ and by
rotation around the same axis through helical step  $\Delta\phi$. These
are the rise and twist parameters introduced in the previous section.
\begin{eqnarray}
x_{n,j,1}&=&x_{n-1,j,1}\cos(\Delta\phi)-x_{n-1,j,2}\sin(\Delta\phi),\nonumber \\
x_{n,j,2}&=&x_{n-1,j,1}\sin(\Delta\phi)-x_{n-1,j,2}\cos(\Delta\phi),\label{f10}\\
x_{n,j,3}&=&x_{n-1,j,3}+\Delta z\nonumber
\end{eqnarray}

Thus, the energy of the ground state is a function of 38 variables:
$\{ {\bf x}_{1,j}\}_{j=1}^{12}$, $\Delta\phi$, $\Delta z$, where
${\bf x}_{1,j}=(x_{1,j,1},x_{1,j,2},x_{1,j,3})$ is the vector position
of  $j$-th grain of the first nucleotide pair.

Finding the ground state amounts to the following minimization problem:
\begin{equation}
E_0=E_v+...+E_{vdW}\rightarrow \min: \{ {\bf x_{1,j}}\}_{j=1}^{12},~\Delta\phi,~\Delta z,
\label{f11}
\end{equation}
where the sum extends over one nucleotide pair $n=1$, and the 
relation (\ref{f10}) holds for
calculation of the energies $E_v$,...,$E_{vdW}$.

Numerical solution of the problem  (\ref{f11}) has shown that the
ground state of poly-G DNA
corresponds to the twist value of $\Delta\phi_0=38.30^\circ$, and the
rise value (z-step) of $\Delta z_0=3.339$\AA.
It should be noticed that if all of the long-range interaction were omitted,
{\it i.e.}, without two last terms $E_q$ and  $E_{vdW}$ in the Hamiltonian (\ref{f1}),
the helical step values would change only slightly, by about 1 per cent:
$\Delta\phi_0=38.03^\circ$, $\Delta z_0= 3.309$\AA.
Thus, long-range electrostatic interactions
between the charged group result in  the relative elongation of the chain
by only  about 1 per cent. Parameters of the double helix computed
within our model differ only slightly from the ``canonical" \ parameters of
the B-conformation of a (heterogeneous)
DNA double helix in the crystal form \cite{Dickerson},
for which the average twist angle is $\Delta\phi = 34^\circ\div 36^\circ$,
and average rise per base pair is $\Delta z = 3.4$\AA .

To find the ground state of the homogeneous double helix under tension,
it is necessary to minimize (\ref{f11}) under the fixed value
of longitudinal step $\Delta z$. As a result,
one can obtain the dependence of the homogeneous state energy on the longitudinal step. 
This function $E_0(\Delta z)$ has  a minimum when $\Delta z = \Delta z_0$, which corresponds
to the B-conformation of the double helix. Longitudinal stiffness of the helix
$K_z = d^2E_0/d\Delta z^2\vert_{\Delta z_0}$. Specifically, within our model
we estimate $K_z=16$ N/m. Since the energy $E_0$ which is being derived is normalized 
to one nucleotide pair one can calculate the stretching modulus 
$S=K_z \Delta z_0 = 16 $ N/m $\times 3.4$ \AA $=5440$pN.
This estimate is somewhat higher than the corresponding estimates 
of $1530 \cdots 3760$ pN
obtained from fluctuations of distances between base pairs observed in MD
simulations\cite{Bruant1999}. The relatively larger value of $K_z$ from our CG model
is consistent with the model's over-all larger stiffness relative to
the all-atom description, see a discussion above.
Some of the difference between the two estimates may also be
due to methodological differences in estimating longitudinal
stiffness. Values of the stretching modulus derived from experiments
are of the order 1000~pN\cite{Cluzel, Smith, Bustamante}, i.e., about 5 times
smaller than our estimate based on the CG model.
One should keep in mind, however, that we have obtained
only an upper estimate for the stretching modulus: temperature
was assumed to be zero, the calculations were based on
a homogeneous poly-G--poly-C sequence that was reported to be more rigid than
inhomogeneous and poly-A--poly-T sequences used in experiments\cite{Clausen-Schaumann2000,Lebrun1996},
and the entropy component was not considered in our calculations.

To obtain $E_0(\Delta\phi)$, that is the dependence of the helix energy on the
helical step $\Delta\phi$, we set $\Delta z\equiv\Delta z_0$ in (\ref{f11})
and perform the minimization with respect to the remaining 36 parameters.
Then, torsion stiffness of the double helix
$K_\phi = \Delta z_0 d^2E_0/d\Delta\phi^2\vert_{\Delta\phi_0}$.
Our estimate, $K_\phi=5.8\times 10^{-28}$ J$\cdot$m, is in
good agreement with the experimental value of
$K_\phi=4.1\pm 0.3 \times 10^{-28}$ J$\cdot$m,
obtained  for DNA macromolecule in B-conformation \cite{Bryant2003}.

For analysis of small-amplitude oscillations of the double helix it is
convenient to use
local cylindrical coordinates ${\bf v}_{n,j}=(v_{n,j,1},v_{n,j,2},v_{n,j,3})$, 
given by the following expressions:
\begin{eqnarray}
x_{n,j,1}&=&x^0_{n,j,1}-v_{n,j,1}\sin\phi_{n,j}+v_{n,j,2}\cos\phi_{n,j}, \nonumber \\
x_{n,j,2}&=&x^0_{n,j,2}+v_{n,j,1}\cos\phi_{n,j}+v_{n,j,2}\sin\phi_{n,j}, \label{f12} \\
x_{n,j,3}&=&x^0_{n,j,3}+v_{n,j,3}, \nonumber
\end{eqnarray}
with ${\bf x}^0_{n,j}$, ($n=0,\pm 1,\pm 2$,...; $j=1$,2,...,12) being
coordinates of the grains in the ground state of the double helix,
and $\phi_{n,j}$ being angular
coordinate of the grain $(n,j)$. Within these new coordinates the molecule's
Hamiltonian (\ref{f1}) has  the following form:
\begin{equation}
H=\sum_n\left[\frac12({\bf M}\dot{\bf v}_n,\dot{\bf v}_n)+P({\bf v}_{n-1},{\bf v}_n,{\bf v}_{n+1})\right],
\label{f13}
\end{equation}
where  ${\bf v}_n=({\bf u}_{n,1},{\bf u}_{n,2},...,{\bf u}_{n,12})$ is a 36-dimensional
vector, ${\bf M}$ is 36-dimensional diagonal mass matrix. Note that the
last two terms $E_q$ and $E_{vdW}$, responsible for
long-range interaction, have been omitted. This simplification is critical
from the methodological point of view, but has very little impact on the
accuracy of the estimates of DNA thermal conductivity. The point will be
discussed below.

Hamiltonian  (\ref{f13}) corresponds to the following system of
equations of motion:
\begin{eqnarray}
-{\bf M}\ddot{\bf v}_n=P_1({\bf v}_{n},{\bf v}_{n+1},{\bf v}_{n+2})\nonumber\\
+P_2({\bf v}_{n-1},{\bf v}_{n},{\bf v}_{n+1})+P_3({\bf v}_{n-2},{\bf v}_{n-1},{\bf v}_{n}),
\label{f14}
\end{eqnarray}
where $P_i({\bf v}_1,{\bf v}_2,{\bf v}_3)=\partial P/\partial{\bf v}_i$, $i=1,2,3$.
Within the linear approximation, the system (\ref{f14}) has  the form
\begin{equation}
-{\bf M}\ddot{\bf v}_n=B_1{\bf v}_n+B_2{\bf v}_{n+1}+B_2^*{\bf v}_{n-1}
+B_3{\bf v}_{n+2}+B_3^*{\bf v}_{n-2}, \label{f15}
\end{equation}
where matrix elements are given by
$$
B_1=P_{11}+P_{22}+P_{33},~~B_2=P_{12}+P_{23},~~B_3=P_{13},
$$
and partial derivative matrix is given by
$$
P_{ij}=\frac{\partial^2P}{\partial{\bf v}_i\partial{\bf v}_j}({\bf 0},{\bf 0},{\bf 0}),~~i,j=1,2,3.
$$

\begin{figure}[tb]
\includegraphics[angle=0, width=1\linewidth]{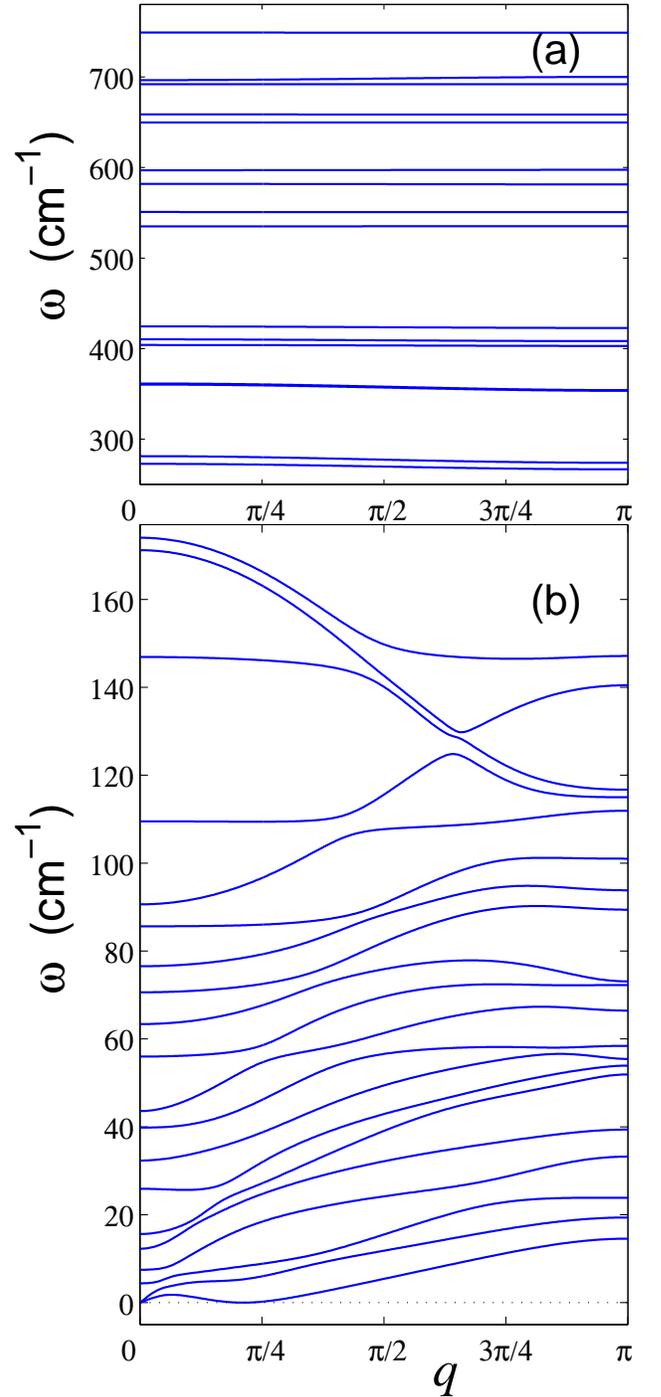}
\caption{\label{fig6}\protect
36 branches of the dispersion curve of homogeneous poly-G DNA:
(a) high-frequency and (b) low-frequency branches.
}
\end{figure}

Solution of the system of linear equations (\ref{f15})
can be found in the standard form
\begin{equation}
{\bf v}_n=A{\bf e}\exp[i(qn-\omega t)], \label{f16}
\end{equation}
where  $A$ is linear mode amplitude, ${\bf e}$ is  unit vector $(|{\bf e}|=1)$,
$q\in [0,\pi]$  is dimensionless wave number. Substituting  
the expression (\ref{f16})
into the system (\ref{f15}), we arrive at the following
36-dimensional eigenvalue problem:
\begin{eqnarray}
\omega^2{\bf M}{\bf e}=[B_1+B_2\exp(iq)+B_2^*\exp(-iq)\nonumber\\
+B_3\exp(2iq)+B_3^*\exp(-2iq)]{\bf e}.\label{f17}
\end{eqnarray}
Thus, to obtain dispersion relations which characterize eigenmodes of the DNA
double helix, 
one has to find all eigenvalues of the problem (\ref{f17}) for each
value of wave number $0\le q\le\pi$.
The calculated dispersion curve includes 36 branches  $\{ \omega_j(q)\}_{j=1}^{36}$ and is
shown on the Fig. \ref{fig6}.

It can be seen from  Fig.~\ref{fig6} that frequency spectrum consists of low-frequency
$0\le\omega\le175$cm$^{-1}$ and high-frequency $\omega\in[267,749]$cm$^{-1}$ domains.
The high-frequency domain describes internal oscillations of the bases.
As  shown in Fig.~\ref{fig6} (a), corresponding dispersion curves have very
small slope,  meaning
that the high-frequency oscillations have a small dispersion.
The low-frequency oscillations have larger dispersion --
see Fig. \ref{fig6} (b). There are two acoustic dispersion curves which include zero
point $(q=0$, $\omega=0)$.  The first curve  $\omega_1(q)$ describes torsional acoustic oscillations,
the second one $\omega_2(q)$ describes longitudinal acoustic oscillations of the double helix.
Thus we can obtain the two sound velocities
$$
v_t=\Delta z\lim_{q\rightarrow 0}\frac{\omega_1(q)}{q},~~
v_l=\Delta z\lim_{q\rightarrow 0}\frac{\omega_2(q)}{q},
$$
with $\Delta z$ being $z$-step of a double helix. The value of the 
torsional sound velocity is
$v_t=850$ m/s, and the
value of the longitudinal sound velocity is $v_l=1790$ m/s. 
One of these dispersion
curves includes the special point $(q=\Delta\phi,\omega=0)$
($\Delta\phi$ is the angular helix step). This curve describes bending oscillations 
of the double helix which we do not analyze in detail because we have so 
far neglected the 
long-range interactions that are known to have strong effect on bending
rigidity of the DNA.

The estimated longitudinal sound velocity is in agreement with 
experimental value of the sound velocity in DNA
fibers \cite{Hakim}:  $v_l=1900$ m/s. Another experimental estimate \cite{Krisch2006} 
of the same quantity is higher, $v_l=2840$ m/s, and was
obtained from inelastic X-ray scattering. The same work reports
torsional sound velocity $v_t=600$ m/s; the 20 \% discrepancy with our
estimate of $v_t=850$ m/s appears acceptable given similar margin of
error seen between different experimental estimates for the longitudinal
velocity.

\section{
Frequency spectrum of the thermal oscillations.
}

Let's again consider a homogenous poly-G DNA chain consisting of $N=200$
base pairs and calculate its frequency spectrum density.
We begin by simulating dynamics of the
helix without taking into account long-range interactions. Later, we will
turn them on to analyze the effect of making this approximation.

To obtain thermalized state of the double helix, the system
of Langevin's equations
(\ref{f8}) should be numerically integrated.
For thermalization of the double helix let's consider initial conditions
corresponding
to the ground state (\ref{f9}), and integrate the system (\ref{f8}) over 
time $t=20t_r$. 
After the equilibration period, the coupling with the thermostat
is switched off,  and the frequency density $p(\omega)$ of the
kinetic energy distribution is obtained. To increase precision,
distribution density was calculated as an average over all grains of the helix.
\begin{figure}[tb]
\includegraphics[angle=0, width=1\linewidth]{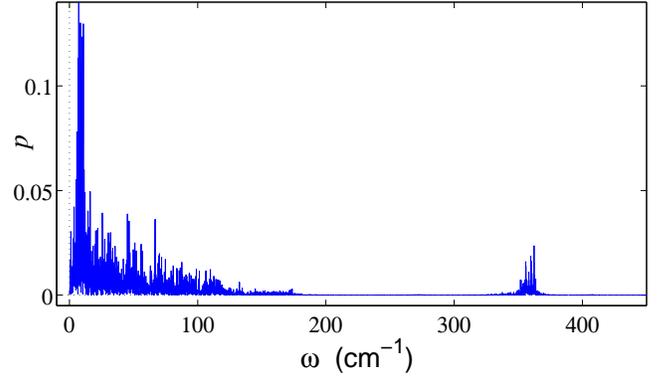}
\caption{\label{fig7}\protect
Frequency spectrum density of the DNA double helix thermal fluctuations at
$T=300$K.
}
\end{figure}

The computed 
frequency spectrum density at $T=300$K is shown in the Fig. \ref{fig7}.
The spectrum is clearly divided into a low-frequency
$0\le\omega\le 175$ cm$^{-1}$  and a high-frequency  $267<\omega<749$ cm$^{-1}$ 
domain, consistent with the dispersion curves of Fig.~\ref{fig6}. 

Simulating the double helix dynamics with account for all  
interactions, including
long-range ones, (results not shown) yields almost the same frequency spectrum. Only the density of
oscillations in the interval   $0\le\omega<10$ cm$^{-1}$ increases somewhat.

\section{
Heat conductivity of the double helix
}
For numerical modeling of the heat transfer along the
DNA double helix, we consider a chain of a fixed length with the
ends placed in two separate thermostats each with its own
temperature.  To calculate the coefficient of thermal
conductivity, we have to calculate numerically the
heat flux through any cross section of the double helix. Therefore,
first we need to obtain a formula for the longitudinal local
heat flux.

Let us consider the homogeneous double helix \linebreak 
poly-G DNA. (The method below is also applicable to any sequences of bases).

If long-range interactions (electrostatic and van der Waals)
are not taken into account we can present the 
Hamiltonian of the helix (\ref{f1})
in the form
\begin{equation}
H=\sum_n\frac12({\bf M}\dot{\bf u}_n,\dot{\bf u}_n)+P({\bf u}_{n-1},{\bf u}_n,{\bf u}_{n+1}),
\label{f18}
\end{equation}
where the first term describes the kinetic energy of atoms in
a given cell and the second term describes the energy of interaction
between the atoms within the cell and with the atoms of  neighboring cells.
The corresponding equations of motion
can be written in the form
\begin{eqnarray}
{\bf M}\ddot{\bf u}_n=-P_1({\bf u}_{n},{\bf u}_{n+1},{\bf u}_{n+2})
                      -P_2({\bf u}_{n-1},{\bf u}_{n},{\bf u}_{n+1})\nonumber\\
                      -P_3({\bf u}_{n-2},{\bf u}_{n-1},{\bf u}_{n}),~~\label{f19}
\end{eqnarray}
where the function $P_j$ is defined as
$$
P_j=\frac{\partial}{\partial{\bf u}_j}P({\bf u}_1,{\bf u}_2,{\bf u}_3),~~j=1,2,3.
$$

To determine the energy flux through the double helix cross
section, we re-write formula (\ref{f18}) in a compact form, $H=\sum_n h_n$, where $h_n$
is the energy density,
\begin{equation}
h_n=\frac12({\bf M}\dot{\bf u}_n,\dot{\bf u}_n)+P({\bf u}_{n-1},{\bf u}_n,{\bf u}_{n+1}).
\label{f20}
\end{equation}

Local longitudinal heat flux $j_n$ is defined through  local
energy density $h_n$ by the discrete version of the continuity
equation,
\begin{equation}
\frac{d}{dt}h_n=j_n-j_{n+1}.
\label{f21}
\end{equation}

Using the energy density (\ref{f20}) and the equations of motion
(\ref{f19}), we can derive the following relations:
\begin{eqnarray*}
\frac{d}{dt}h_n=({\bf M}\ddot{\bf u}_n,\dot{\bf u}_n)+(P_{1,n},\dot{\bf u}_{n-1})
+(P_{2,n},\dot{\bf u}_n)\\
+(P_{3,n},\dot{\bf u}_{n+1})=-(P_{1,n+1},\dot{\bf u}_n)-(P_{3,n-1},\dot{\bf u}_n)\\
+(P_{1,n},\dot{\bf u}_{n-1})+(P_{3,n},\dot{\bf u}_{n+1}),
\end{eqnarray*}
where
$$
P_{j,n}=P_j({\bf u}_{n-1},{\bf u}_{n},{\bf u}_{n+1}),~~j=1,2,3.
$$
From this and (\ref{f21}) it follows that the energy
flux through the $n$-th cross section has the following simple
form:
\begin{equation}
j_n=(P_{1,n},\dot{\bf u}_{n-1})-(P_{3,n-1},\dot{\bf u}_n).
\label{f22}
\end{equation}
Let us note that taking into account long-range interactions
would complicate this formula considerably, making the calculations
virtually intractable. This is why the approximation we have made is 
critical.

For a direct numerical modeling of the heat transfer along
the double helix, we consider a finite structure of the
length $N\Delta z$ with fixed ends. We assume that the first $N_+=20$ segments are
placed in the thermostat at temperature $T_+=310$ K and
the last $N_-=20$ segments are placed in the other thermostat at
$T_-=290$ K. The helix dynamics is described by the following
equations of motion:
\begin{eqnarray}
{\bf M}\ddot{\bf u}_n&=&-{\bf F}_n -\Gamma{\bf M}\dot{\bf u}_n+\Xi_n^+,~~n=1,...,N_+,\nonumber\\
{\bf M}\ddot{\bf u}_n&=&-{\bf F}_n,~~n=N_++1,...,N-N_-, \label{f23}\\
{\bf M}\ddot{\bf u}_n&=&-{\bf F}_n -\Gamma{\bf M}\dot{\bf u}_n+\Xi_n^-,~~n=N-N_-+1,...,N,
\nonumber
\end{eqnarray}
where ${\bf F}_n=\partial H/\partial{\bf u}_n$,
$\Gamma=1/t_r$ is the damping coefficient (relaxation time $t_r=1$ ps,
and  $\Xi_n^\pm=(\xi_1^\pm,...,\xi_{36}^\pm)$
is a 36-dimensional vector of normally distributed
random forces normalized by the condition
$$
\langle \xi_{n,i}^\pm(t_1)\xi_{m,j}^\pm(t_2)\rangle=2Mk_BT_\pm\delta_{nm}\delta_{ij}\delta(t_2-t_1),
$$
where the mass $M=M_k$, if $i=3(k-1)+l$, $k=1,...,12$, $l=1,2,3$.

We take the initial conditions (\ref{f9}) corresponding to the equilibrium
state of the helix.
With these initial conditions, we integrate
the equations of motion (\ref{f23}) numerically, by employing
the velocity Verlet method with step $\Delta t=0.0005$~ps.
After integration time $t_0$ [this value depends on the
helix length between the thermostats, $\Delta L=(N-N_+-N_-)\Delta z$],
we observe the formation of a temperature gradient
and a constant heat energy flux in the central part of the helix.
It is important to notice that the time $t_0$ can be reduced by modifying the
initial distribution of the energy, {\it e.g.}, by taking the initial condition for the system
(\ref{f23}) as homogeneously thermalized state with the
mean temperature  $T=(T_++T_-)/2=300$~K.

After the stationary heat flux is established,
the temperature distribution can be found  using the formula
$$
T_n=\lim_{t\rightarrow\infty}\frac{1}{36k_Bt}\int_0^t({\bf M}\dot{\bf u}_n(\tau),\dot{\bf u}_n(\tau))d\tau
$$
and the averaged value of the energy flux along the helix
$$
J_n=\lim_{t\rightarrow\infty}\frac{\Delta z}{t}\int_0^tj_n(\tau)d\tau.
$$
\begin{figure}[tb]
\includegraphics[angle=0, width=1\linewidth]{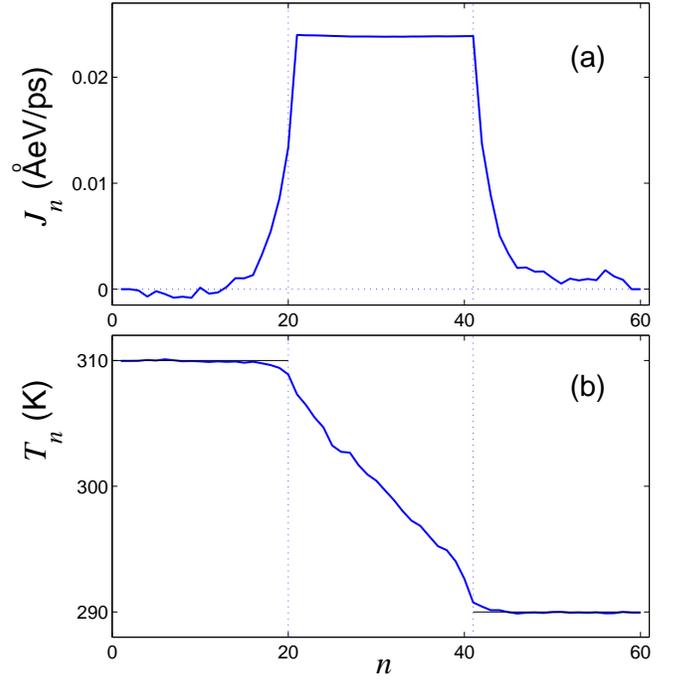}
\caption{\label{fig8}\protect
Distributions of (a) local heat flux $J_n$ and
(b) local temperature $T_n$ in the double helix
with length $N\Delta z$. The input parameters are $N=60$, temperatures
$T_+=310$ K and $T_-=290$ K, and the number of cells in the
thermostats, $N_\pm=20$.
}
\end{figure}

Distributions of the local energy flux and temperature
along the helix are shown in Figs. \ref{fig8} (a) and (b). In the
steady-state regime, the heat flux through each of the cross
section at the central part of the helix should remain the
same, {\it i.e.}  $J_n\equiv J$, $N_+<n\le N-N_-$. This property can be employed
as a criterion for the accuracy of numerical modeling
and can also be used to determine the characteristic time for
achieving the steady-state regime and calculation of $J_n$ and
$T_n$. Figure \ref{fig8} (a) suggests that the flux is constant along the
central part of the helix indicating that we have reached the
required regime.

At the central part of the helix, we observe a linear
gradient of the temperature distribution, so that we can define the coefficient
of thermal conductivity as
\begin{equation}
\kappa(N-N_+-N_-)=\frac{(N-N_--N_+-1)J}{(T_{N_++1}-T_{N-N_-})S},
\label{f24}
\end{equation}
where $S=\pi R^2$ is the area of the cross section of the double helix
($R=8$ \AA~ is the radius of helix on phosphorus atoms).
In this way, the calculation of thermal conductivity is reduced to the calculation
of the limiting value,
$$
\kappa=\lim_{N\rightarrow\infty} \kappa(N).
$$

In order to determine the coefficient of thermal conductivity,
we need to know only the dependence of the temperature
from base-pair number in the central part of the helix. However, a change of
the temperature distribution at the edges of the helix can
also provide some useful information. If the helix is
placed into a Langevin thermostat at temperature $T$, each
segment of the helix should have the temperature $T_n=T$
due to the energy balance of the input energy from random
forces and the energy lost to dissipation. Then, an averaged
energy flow from the $n$-th segment of the helix can be
presented as
$$
\Gamma({\bf M}\dot{\bf u}_n,\dot{\bf u}_n)=36k_BT_n/t_r.
$$
If only the edges of the helix are placed into thermostat,
there appears an additional energy exchange with its central
part, so the energy from the right edge will flow to the left
one. As a result, the temperature of the left edge is reduced
$(T_n\le T_+$, $n=1,2,...,N_+)$, whereas the temperature at the right edge increases
$(T_n\ge T_-$, $n=N-N_-+1,...,N)$ -- see Fig. \ref{fig8} (b).
This information allows us to find the
energy flux in the central part of the double helix  using only the
energy imbalance at the edges,
\begin{equation}
\frac{Jt_r}{\Delta z 36 k_B}=\sum_{n=1}^{N_+}(T_+-T_n)=\sum_{n=N-N_-+1}^{N}(T_n-T_-).
\label{f25}
\end{equation}
If the lengths of the edges placed into thermostat coincide,
i.e., $N_+=N_-=N_\pm$, we can rewrite this formula in the following
simplified form:
\begin{equation}
J=\frac{\Delta z18k_B}{t_r}\sum_{n=1}^{N_\pm}(T_+-T_--T_n+T_{N+1-n}).
\label{f26}
\end{equation}

Equation (\ref{f25}) gives an alternative way to calculate 
thermal energy flux $J$; the equation can be employed to verify
results obtained via Eq. (\ref{f22}).
Let us note that although (\ref{f22}) is obtained under
the assumption of no long-range
interactions, formula (\ref{f26}) remains valid also if these interactions
are taken into account.

Numerical modeling of the heat transfer shows that both formulas lead to the same
value of the heat-conductivity coefficient if long-range interactions are absent.
When N=80 (the number of internal links $N_i=N-N_+-N_-=20$), the heat-conductivity
coefficient $\kappa=0.26$ W/mK. When $N=80$ ($N_i=40$) -- conductivity $\kappa=0.29$ W/mK,
when $N=120$ ($N_i=80$) -- $\kappa=0.27$ W/mK, and
when $N=200$ ($N_i=160$) -- $\kappa=0.28$ W/mK.
The same values are obtained also if the
long-range interactions are taken into account (and the heat flow is calculated
by formula (\ref{f26}) only). These considerations help us reach
the conclusion that the contribution of the long-range interactions
to the heat transfer along the double helix is very minor.

It is worth noting  that the use of formula (\ref{f26}) for calculating
the value of heat transfer requires more time-consuming calculations.
Therefore, it is preferable to use formula (\ref{f22}).
Also, equation (\ref{f22}) allows one to estimate relative 
contributions of various 
interactions into the process of heat transfer. We find that interaction 
between neighboring 
base pairs  contributes 32\% to the net energy flow, with 
the  rest of the heat transfer occurring along the two sugar-phosphate chains.

As one can see from the results, the value of heat conductivity $\kappa$
in the DNA macromolecule does not depend on the length of the molecule. This 
is normal thermal conductivity for which Fourier's law is valid 
at nano-level as well, at least as far as the DNA is concerned.
This is in contrast to earlier models of heat conduction along carbon nanotubes
and nanoribbons that predicted anomalous thermal conductivity -- 
divergence of the coefficient of thermal conductivity with sample
length\cite{chang2,maruyama,zhang,sav1}. Compared to nanotubes, the DNA 
double helix is much softer, which leads to 
strongly nonlinear behavior at $T=300$ K (in contrast, a nanotube is 
a rigid quasi-one-dimensional structure, with only weak nonlinear dynamics).
Contribution of nonlinearity to the DNA dynamics will be explored in more
detail in the following section. 

\section{Dependence of the thermal conductivity on temperature}

\begin{figure}[tb]
\includegraphics[angle=0, width=1\linewidth]{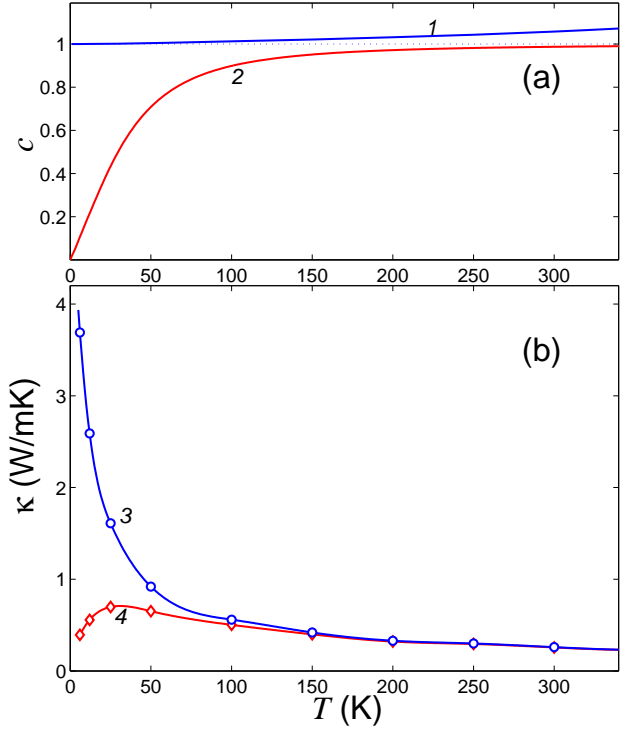}
\caption{\label{fig9}\protect
(a) Temperature dependence of dimensionless specific heat $c(T)$ and $c_q(T)$ 
(curves 1 and 2, respectively);
b) heat conductivity $\kappa(T)$ and $\kappa_q(T)$ (curves 3 and 4, respectively) 
of the DNA double helix.
The dependencies $c(T)$ and $\kappa(T)$ are obtained in the 
framework of classical molecular dynamics model, while $c_q(T)$ and $\kappa_q(T)$ are computed 
within the quantum framework. 
}
\end{figure}

At $T=300$ K the DNA double helix exhibits high-amplitude vibrations 
(the amplitudes can be estimated from Fig. \ref{fig4} and \ref{fig5}). 
The contribution of nonlinearity to the DNA 
dynamics can be estimated from the temperature dependence
of dimensionless heat capacity
\begin{equation}
c(T)=\frac{1}{36Nk_BT}\frac{d}{dT}E(T),
\label{f28}
\end{equation}
where $E(T)=\langle H\rangle$ is average double helix energy at temperature $T$.
For a harmonic system, 
dimensionless heat capacity $c(T)\equiv 1$; for a system
with strong anharmonism $c(T)<1$, and $c(T)>1$ for weakly anharmonic systems.
As seen from Fig.~\ref{fig9}, heat capacity of the 
double helix equals to 1 for low
temperatures ($T<10$ K) and increases  monotonously when the temperature grows.
The heat capacity $c=1.05$ at $T=300$ K, implying weak anharmonism. 

The role of nonlinearity decreases monotonously as the temperature decreases. 
In the limiting case $T\rightarrow 0$ the double helix becomes harmonic. 
Therefore, classical thermal conductivity has to increase monotonously 
as the temperature 
decreases, and diverge when $T\rightarrow 0$. The results of our 
numerical modeling 
confirm  this conclusion -- see Fig.~\ref{fig9} (b), curve 3. 
At  $T\searrow 0$ the heat conductivity  $\kappa\nearrow\infty$.

We should mention that the temperature dependence of the DNA 
thermal conductivity
found above is obtained with the framework of classical molecular-dynamics model, which
does not take into account quantum effects of "frozen" \ high-frequency oscillations
(to take those into account requires substantial modifications to the 
model\cite{bsh2008,dclhg2009}). In crystals at low temperatures,
thermal conductivity decays monotonically when $T\rightarrow 0$.
This is explained by the fact that at low temperatures the temperature dependence 
of thermal conductivity is defined mainly by the temperature dependence
of heat capacity.
 
In classical mechanics, heat capacity of phonons does not depend on temperature,
whereas in quantum mechanics such a dependence is defined by the formula
$c(\omega,T)=k_BF_E(\omega,T)$, where the Einstein function
$$
F_E(\omega,T)=\left(\frac{\hbar\omega}{k_BT}\right)^2
\frac{\exp(\hbar\omega/k_BT)}{[\exp(\hbar\omega/k_BT)-1]^2},
$$
where $\omega$ is the phonon frequency ($0\le F_E\le 1$, 
function $F_E\searrow 0$ for $T\searrow 0$ and $F_E\nearrow 1$ for $T\nearrow\infty$).

As seen from the DNA dispersion curves $\{\omega_i(q)\}_{i=1}^{36}$, the 
main contribution in the heat conductivity is determined by the 
20 low-frequencies phonons 
(16 high-frequencies phonons have very small group velocities, and therefore can not 
be efficient energy carriers). The temperature dependence of dimensionless heat capacity 
of low frequencies phonons can be found using formula
\begin{equation}
c_q(T)=\frac{1}{20\pi}\sum_{i=1}^{20}\int_0^\pi F_E(\omega_i(q),T)dq.
\label{f29}
\end{equation}
One can see from Fig.~\ref{fig9} that the heat capacity  $c_q$ does not noticeably depend
on temperature if $T>150$~K, and tends monotonously 
to zero as the temperatures decrease below $T<150$~K. 

Thus, thermal vibrations of the double helix can be described 
classically for $T>150$~K only. 
For lower temperatures, quantum effect caused by "freezing out" 
of high-frequency
vibrations must be taken into account. Due to these effects 
the DNA heat capacity  (\ref{f29}) 
tends monotonously to zero as the temperature decreases. 
The double helix thermal conductivity $\kappa_q(T)\approx c_q(T)\kappa(T)$, 
(where the temperature dependence $\kappa(T)$ is calculated classically) because
the phonon energy is proportional to heat capacity.
As it seen form Fig.~\ref{fig9} (b) 
at $T>30$~K the thermal conductivity $\kappa_q$ grows monotonously as 
the temperature decreases, reaching its maximum 
at $T\approx 30$~K, and then decreases monotonously as $T \rightarrow 0$.

These calculations show that heat transfer in the DNA 
occurs mainly due to propagation 
of low-frequency phonons (frequencies $\omega<175$~cm$^{-1}$), i.e., by ``soft" 
low-frequencies waves. Such oscillations are strongly coupled to deformation 
of orientation angles. This fact clearly distinguishes the 
DNA double helix from the essentially rigid 
carbon nanotubes and nanoribbons. The simplest model of a 
one-dimensional system with 
orientational interaction is one-dimensional chain of interacting rotators.
This chain has a finite thermal conductivity \cite{Gend00,Giardina}. 
On the other hand,
nanotubes and nanoribbons are commonly described in 
the one-dimensional approximation as anharmonic Fermi Pasta Ulam (FPU)  
chains that lead to infinite heat conductivity \cite{Lepri97,Lepri03}. 

Thus, the double helix of a homogeneous poly-G DNA has a finite
thermal conductivity $\kappa=0.3$ W/mK. The 
double helix with a nonhomogeneous (arbitrary) base sequence
may be expected to have a smaller value of the heat conductivity coefficient since the presence
of inhomogeneities leads to additional phonon scattering. Therefore,
thermal conductivity of a generic DNA double helix,
$\kappa\le 0.3$ W/mK, may be expected to be less than half of that  of  water
heat conductivity which is 0.6 W/mK. This means that DNA macromolecule
is a thermal insulator relative to its surrounding solution.
It should be noted that
experimentally measured  
thermal conductivity of the DNA-gold composite structure (DNA is
a matrix for gold nano-particles) \cite{kodama} gives the coefficient of thermal conductivity
150 W/mK,  which is 500 times higher than the predicted
thermal conductivity of pure DNA. Thus, we conclude
that the measured thermal conductivity of the
DNA-gold composite is completely determined  by the metal component, not the
DNA.

\section{Conclusions}
A coarse-grain (12CG) model of DNA double helix is proposed in which each nucleotide is
represented by 6 "grains". The corresponding effective pair potentials are inferred from
correlation functions obtained from classical all-atom molecular dynamics (MD) trajectories and
potentials (AMBER). The computed structural characteristics  
and fluctuations of the double helix at $T=300$ K are in reasonable agreement with  
available experimental data and earlier computations based on all-atom models. 
An analysis of dispersion curves derived from the coarse-grained model yields 
longitudinal and torsional sound velocities in close agreement with  experiment.

The numerical modeling of heat conductivity along a single DNA molecule
shows that double DNA helix has a finite (normal) thermal conductivity. This means 
that Fourier's law is valid at nano-level for the DNA, 
i.e., coefficient of thermal conductivity does not depend on the length 
of the DNA fragment. Single molecule DNA thermal conductivity 
does not exceed 0.3~W/mK, which is
two times smaller than thermal conductivity of water. Thus, 
DNA double helix is a poor heat conductor. At the same time, it is known from modeling 
of heat transfer along carbon nanotubes and nanoribbons
that the coefficient of thermal conductivity in these systems 
diverges as the specimen length grows \cite{chang2,maruyama,zhang,sav1}.
The anomalous behavior of thermal conductivity in  long nano-objects is
caused by their rigid structure as well as by their weakly nonlinear
quasi one-dimensional
dynamics, mostly due to rigid covalent interactions. In contrast,
the DNA double-helix is a soft 3D structure with strongly
nonlinear dynamics. 
Based on the results of our coarse-grained simulations we conjecture
that heat conduction along the double helix is due predominantly
to weak non-valent orientational interactions.

\section{Acknowledgements}

This research was supported by RFBR (grant 08-04-91118-a) and CRDF (grant RUB2-2920-MO-07).
The authors also thank the Joint Supercomputer Center of the Russian Academy of Sciences
for access to their computer facilities.

\appendix 
\section{Masses of the coarse grains.} 
The mass of each of the backbone grains [P],[C3] and [C1] is calculated as a 
sum of the masses of the original atoms 
included in the grain, Fig. \ref{fig2}. 
So $m_{[P]}=109$ a.e., $m_{[C3]}=26$ a.e., $m_{[C1]}=43$ a.e. 
The distribution of the total mass of base X (X = A, T, G, C) 
between its three defining grains, $m_1 , m_2 , m_3$, can be found
from the condition of preserving the total mass  and preserving the position of the center of mass of the base.
Values of the grain masses are shown in table \ref{tab1}.

\begin{table}[t]
\caption{Masses of the three coarse 
grains ($m_1, m_2, m_3$) for each of the base
X$=$A, T, G, C. In units
of proton mass $m_p$ }
\label{tab1}
\begin{tabular}{cccc}
\hline
X &  ~~$m_1$  & ~$m_2$ & ~$m_3$  \\
\hline
A & ~~$52.230~$ & ~$28.139$~ & ~$53.632$~ \\
T & $51.822$ & $16.204$ & $56.974$ \\
G & $61.731$ & $34.357$ & $53.912$ \\
C & $39.254$ & $35.492$ & $35.254$ \\
\hline
\end{tabular}
\end{table}

\section{The potential function.}
For convenience let's re-write the Hamiltonian of the system:
\begin{equation}
H=E_k+E_v+E_b+E_a+E_t+E_{hb}+E_{st}+E_{el}+E_{vdW}.
\label{f1}
\end{equation}
The first term is the kinetic energy of the system:
\begin{equation}
E_k=\sum_{n=1}^{12N} \frac12M_i\dot{\bf r}_{i}^2~, \label{f2}
\end{equation}     
where the summation is over all $12N$ coarse-grain particles (grains) 
in the system.
\begin{figure}[h]
\includegraphics[angle=0, width=\linewidth]{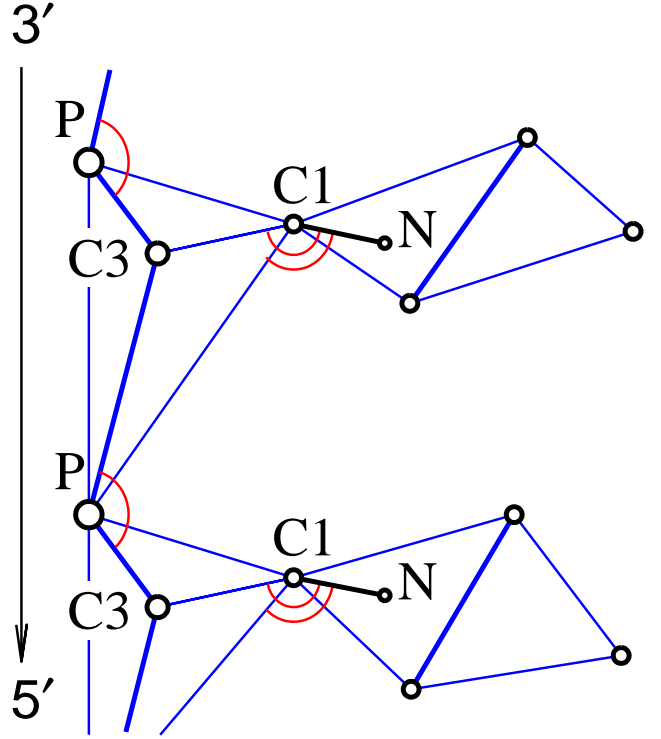}
\caption{\label{fig10}\protect
Grains involved in valent interactions. Blue lines denote valent (harmonic) bonds, 
red arcs mark valent angles, bold blue lines are axes of rotation 
in the torsion potentials. 
The circles marked as N stand for original 
atoms N9 on A,G bases and N1 on T,C bases 
(no coarse grains are centered on these atoms, their coordinates 
are calculated from the positions of the three grains that define the base
plane).
}
\end{figure}

The second term $E_v$ in the Hamiltonian (\ref{f1})
stands for deformation energy of \ "valence" \ (pair) bonds.
Pair potentials have the standard form 
\begin{equation}
U_{\alpha \beta}({\bf x}_1,{\bf x}_2)
=\frac12 K_{\alpha \beta}(|{\bf x}_2-{\bf x}_1|-R_{\alpha \beta})^2, \label{f4}
\end{equation}
where  $\alpha \beta$ denotes types of bonded particles (for example, P and C3), 
parameter $R_{\alpha \beta}$
is the equilibrium length, parameter $K_{\alpha \beta}$  is the  
bond stiffness. Values of these 
parameters were obtained by analysis of all-atomic MD trajectories. 
These potentials are 
calculated for the following pairs: P and C3, C3 and C1, C3 and P, P and C1, C1 and P, P and P
(from neighbouring sites). Order in a pair corresponds to direction from 3'-end to 5'-end 
(see Fig. \ref{fig10}).  
The parameter values are given in the table \ref{tab2}.
\begin{table}[t]
\caption{
Values of the stiffness coefficients $K_{\alpha \beta}$ and bond lengths $R_{\alpha \beta}$ 
for pair interaction potentials
$U_{\alpha \beta}({\bf x}_1,{\bf x}_2)$.
}
\label{tab2}
\begin{tabular}{ccccccc}
\hline
$\alpha \beta$             &  PC3   & C3C1   &  C3P   &  PC1   &  C1P   & PP  \\
\hline
$K_{\alpha \beta}$ (eV/\AA$^2$) & 9.11   & 8.33   & 0.694  & 0.66   & 0.781  & 0.20 \\
$R_{\alpha \beta}$ (\AA)        & 2.6092 & 2.3657 & 4.0735 & 3.6745 & 4.8938 & 6.4612\\
\hline
\end{tabular}
\end{table}

The third term $E_b$ in the Hamiltonian (\ref{f1}) describes base deformation energy. 
This term was introduced to keep all four points near one plane and serves 
to mimic valent 
interaction in nitrogen bases. Let's denote the position of C1 particle by ${\bf x}_1$ 
and positions of the three particles on a base by ${\bf x}_2, {\bf x}_3, {\bf x}_4$. The deformation 
energy includes harmonic constraints on pair distances and a constraint 
on the bending angle of the rectangle \{ ${\bf x}_1$ ${\bf x}_2, {\bf x}_3, {\bf x} _4$ \} 
around its diagonal. Thus, base $\gamma$ ($\gamma$=A, T, G, C) 
deformation energy is given by the following formula:
\begin{eqnarray}
U_{\gamma}({\bf x}_{1},{\bf x}_{2},{\bf x}_{3},{\bf x}_{4})=
 \frac12K_\gamma\left[ (|{\bf x}_1-{\bf x}_2|-R_{\gamma 12})^2~ \right.\nonumber\\
+(|{\bf x}_1-{\bf x}_4|-R_{\gamma 14})^2
+(|{\bf x}_2-{\bf x}_3|-R_{\gamma 23})^2~~\nonumber\\
+(|{\bf x}_2-{\bf x}_4|-R_{\gamma 24})^2
\left.+(|{\bf x}_3-{\bf x}_4|-R_{\gamma 34})^2\right]\nonumber\\
+\epsilon_{\gamma}(1+\cos\theta), \label{f6}
\end{eqnarray}
where $\theta$ is the angle between the 
two planes  ${\bf x}_1{\bf x}_2{\bf x}_4$ and
${\bf x}_2{\bf x}_3{\bf x}_4$ (equilibrium corresponds to all four points lying
on one plane and $\theta=\pi$).
The values of potential parameters can be found in table \ref{tab3}.
Parameters $R_{\gamma 14}$,...,$R_{\gamma 34}$ were defined as equilibrium distances between corresponding
points on bases, values of parameters $K_\gamma$ and $\epsilon_\gamma$ were determined from analysis
of frequency spectrum of base oscillations in all atomic DNA molecular dynamics \cite{amb1}.

\begin{table}[b]
\caption{Values of parameters for potential $U_{X}$ describing
deformation of the base X=A, T, G, C.}
\label{tab3}
\begin{tabular}{lcccc}
\hline
$\gamma$ & A & T & G & C \\
\hline
$R_{\gamma 12}$ (\AA) & ~2.6326~ & ~5.0291~ & ~2.5932~ & ~2.4826~ \\
$R_{\gamma 14}$ (\AA) & 4.3195 & 2.7007 & 5.2651 & 2.6896 \\
$R_{\gamma 23}$ (\AA) & 4.2794 & 2.8651 & 4.2912 & 3.5882 \\
$R_{\gamma 24}$ (\AA) & 4.3111 & 5.5150 & 5.6654 & 3.5014 \\
$R_{\gamma 34}$ (\AA) & 3.5187 & 4.5399 & 4.5807 & 4.5523 \\
$K_\gamma$ (eV/\AA$^2$) & 30 & 30     & 30     & 20     \\
$\epsilon_{\gamma}$ (eV) & 100& 100    & 150    & 70     \\
\hline
\end{tabular}
\end{table}

The fourth term $E_a$ in the Hamiltonian \ref{f1}
describes the energy of angle deformation  and has following form:
$$
U_a(\theta)=\epsilon_a(\cos\theta-\cos\theta_a)^2,
$$

This energy is calculated for following angles:
C3-P-C3, C3-C1-N, N-C1-P. Here N denotes a specific nitrogen atom 
atom on the base: atom N9 for bases 
A and G, and atom N1 for bases T and C. Equilibrium angle and deformation energy 
are summarized in the table \ref{tab_new1}.

\begin{table}[tb]
\caption{
Values of deformation energy $\epsilon_X$ and equilibrium angle $\theta_X$ for angle potentials.
}
\label{tab_new1}
\begin{tabular}{cccc}
\hline
$type$             &  C3-P-C3   & C3-C1-N   &  N-C1-P  \\
\hline
$\epsilon_a$ (eV) &  $0.5$ & $3.$      & $0.3$        \\
$\theta_a$      &  $130.15^\circ$  &   $141.63^\circ $     &      $87.17^\circ $  \\
\hline
\end{tabular}
\end{table}

The fifth term $E_t$ in the Hamiltonian (\ref{f1}) describes torsional 
deformation energy. 
It has the form:
$$
U_{t}=\epsilon_{t}(1-\cos(\phi-\phi_0))
$$

The first type of potential is for the torsion C3-C1-N9-C8 (C3-C1-N1-C6) -- i.e., rotations of  base
A, G (T, C) around the bond  C1---N9 (C1--N1).
The second type of potential is for the torsion C3-P-C3-C1, the third one for the torsion C1-C3-P-C3.
Parameters of these potentials are summarized in table \ref{tab_new2}.

\begin{table}[tb]
\caption{
Deformation energy $\epsilon_{t}$ and equilibrium values $\phi_0$ for 
the torsional potentials.
}
\label{tab_new2}
\begin{tabular}{cccc}
\hline
Potential         &  C3-C1-N-C   & C3-P-C3-C1   &  C1-C3-P-C3  \\
\hline
$\epsilon_{t}$ (eV) &  $0.5$ & $0.5$      & $0.5$        \\
$\phi_0$      &  $0$  &   $-26.21^\circ $   & $48.58^\circ $  \\
\hline
\end{tabular}
\end{table}

The sixth term $E_{hb}$ in the Hamiltonian (\ref{f1}) describes the
energy of interaction between complementary bases.
Since each nitrogen base is a rigid planar structure, one can restore positions
of all of its original atoms from positions of the three coarse-grain atoms,
as outlined in the previous section. Let's denote the 
set of coordinates of three coarse-grain 
atoms by $X_n$ with $n$ being a number of the base-pair. 
One can calculate coordinates of all of the original atoms 
on the base: $r_1(X_n), r_2(X_n),\dots$. Hence we can use the proven  
all-atom AMBER 
(van der Waals and electrostatics) potentials \cite{amb1} for hydrogen bonds
and stacking interactions.
Thus
$$
E_{hb}=\sum_n V_{XY}(X_n,Y_n)=$$
$$\sum_n U_{AMBER}(r_1(X_n), r_2(X_n),\dots,r_1(Y_n), r_2(Y_n),\dots).
$$
where $V_{XY}(X_n,Y_n)$ is a
potential of interaction between X (X=A,T,G,C) base and complementary Y (Y=A,T,G,C) base.

\begin{figure}[tb]
\includegraphics[angle=0, width=1\linewidth]{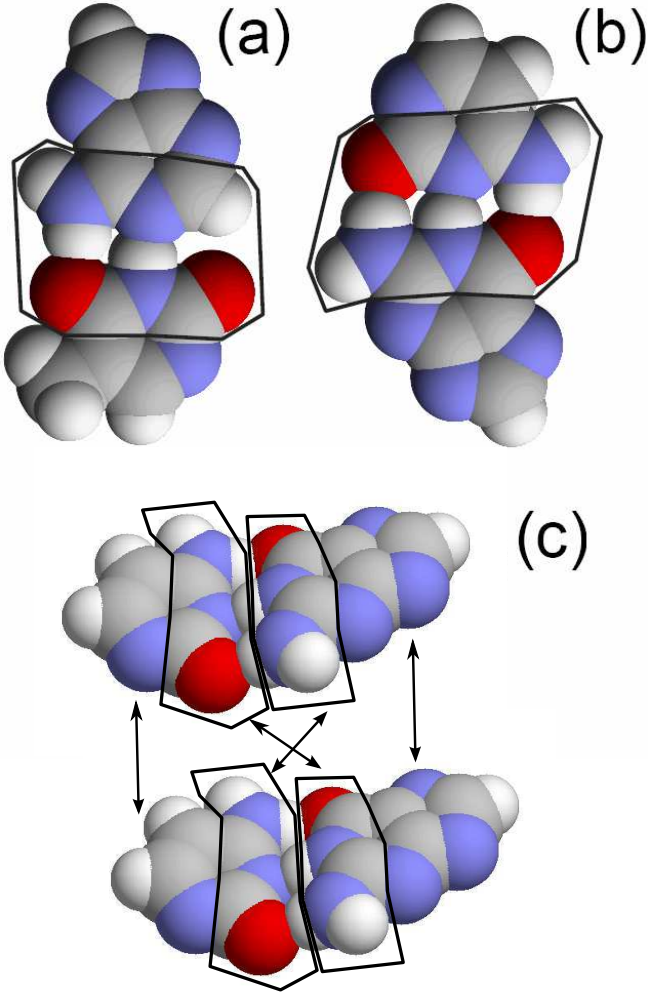}
\caption{\label{fig11}\protect
View of (a) AT base pair, (b)  GC base pair
(highlighted are atoms which contribute most to base-base interaction energy)
and (c) two neighboring base-pairs (AT and GC). Arrows indicate parts of nitrogen bases 
whose interaction is taken into account: for bases on the complementary 
strands only those atoms that face 
another contribute to the interaction, while for neighboring bases on 
the same strand all of the atoms contribute.
}
\end{figure}

The main part of the hydrogen bond energy is interactions between atoms
near the hydrogen bond -- see Fig. \ref{fig11} (a) and (b).
Hence the number of interacting atoms can be reduced. 
Let's denote this ``reduced" potential by
$V^*_{XY}(X_n,Y_n)$.
Then
$$
E_{hb}=\sum_n V^*_{XY}(X_n,Y_n).
$$

The interaction energy between neighboring bases is given by
$$
E_{st}=\sum_n
V_{XY}(X_n,X_{n+1})+V^*_{XY}(X_n,Y_{n+1})+
$$
$$
V_{XY}(Y_n, Y_{n+1})+V^*_{XY}(Y_n, X_{n+1}).
$$

Atoms whose interactions are taken into account in calculation of 
the interaction energy between neighbor bases 
are shown in Fig. \ref{fig11} (c).

The eighth term $E_{el}$ of the Hamiltonian  (\ref{f1}) describes
the charge-charge interactions within the double helix. Within our model,
only the phosphate groups interact via long-range electrostatic forces.
We assume that each [P] grain carries 
charge equal to the electron charge $q_P=-1e$,
while all other particles are neutral.
The total electrostatic energy of the DNA
in aqueous environment (including ions) is
written as $E_{el} = E_{vac}  +  \Delta G_{solv}$, where $E_{vac}$
represents the Coulomb interaction energy in vacuum,
and $\Delta G_{solv}$ is defined as the free energy of transferring the
molecule from vacuum  into solvent, {\it i.e.},  solvation free energy.
The above decomposition is an
approximation made by most classical (non-polarizable) potential. Within
our model we further
assume that $\Delta G_{solv}$ contains only the electrostatic part;
this is a reasonable assumption as long as the shape of the DNA double-helix
does not change drastically
during dynamics (e.g., the strands do not separate), and thus changes
in the "hydrophobic" part of $\Delta G_{solv}$ can be neglected. While
computation of the Coulomb part of the interaction is trivial, estimation
of $\Delta G_{solv}$ is not, due to non-trivial shape of the biomolecule. Within
the framework of the continuum dielectric, linear response theory the principle
way of estimating $\Delta G_{solv}$ is by  
solving the Poisson-Boltzmann (PB) equation
with the boundary conditions determined by the molecular surface that separates the high
dielectric solvent from the low dielectric interior of the molecule.
However, the corresponding procedures are expensive, and currently of
limited practical use in dynamical simulations. We therefore resort to
the so-called generalized Born model \cite{Still1990,Tsui-Case,Bashford}
(GB), which is the most widely used alternative to the PB treatment when speed of computation
is a concern, particularly in
molecular dynamics \cite{OnufrievAnnRep08}, including simulations of
nucleic acids \cite{Chocholousova06,Keslo06,Zacharias06,Wang02,Tsui2001,sorin03,Balaeff:1998:Proteins:9489920,Jayaram:2002:J-Comput-Chem:11913374,DeCastro:2002:J-Mol-Recognit:12382239,Allawi03,Onufriev1}.

The GB model approximates $\Delta G_{solv}$ by the following formula
proposed by Still et al. \cite{Still1990}
\begin{equation} \label{eqn:DGgb}
  \Delta G_{solv} \approx
   - \frac{1}{2} \left( 1 - {{1}\over{{\epsilon}_{out}}} \right)
   \sum_{ij} \frac{q_i q_j}{f(r_{ij}, R_i, R_j)},
\end{equation}
where $\epsilon_{out}$ is the dielectric constant of water,
$r_{ij}$ is the distance between atoms $i$ and $j$,
$q_i$ is the partial charge of atom $i$,
$R_i$ is the so-called \emph{effective Born radius} of atom $i$,
and $f = {\Big[r_{ij}^2 + R_i R_j \exp({-r_{ij}^2/{4 R_i R_j})} \Big]}^{1\over2}$.
The empirical function is designed to interpolate between the limits
of large $r_{ij} \gg \sqrt{R_i R_j}$  where the Coulomb law applies, and
the opposite limit where the two atomic spheres fuse into one, restoring the
famous Born formula for solvation energy of a single ion.
The effective Born radius of an atom represents its degree of burial
within the low dielectric interior of the molecule: the further away is
the atom from the solvent, the larger is its effective radius.
In our model, we assume constant effective Born radii which we calculate
once from the first principles \cite{Onufriev1102}. The screening effects
of monovalent salt are introduced approximately, at the Debye-Huckel level by
substitution
$$
1 - {\epsilon_{out}}^{-1} \rightarrow
1 - {\epsilon_{out}}^{-1} \exp( -0.73 \kappa f).
$$
The 0.73 pre-factor was found empirically to give the best agreement with the numerical
PB treatment \cite{Case1999}. Here $\kappa$ is the Debye-Huckel screening parameter
$\kappa$[\AA$^{-1}$]$\approx 0.316 \sqrt{{\rm[salt][mol/L]}}$.

Further simplifications come from the fact that we have only one non-zero
charge species in our model, the [P] grain.
Then, the total  electrostatics energy is given by
\begin{eqnarray*}
E_{el}=C_0+\sum_{i,j=1}^{N_P} V_q(r_{ij})
\end{eqnarray*}
where the summation is performed over all different [P]-grains pairs  
where
\begin{equation}\label{f7}
V_q(r)=
C_1 \left[\frac{1}{r}-\frac{1}{f(r)}\left(1-\epsilon_{out}^{-1}e^{-0.73 \kappa f(r)}\right)\right]
\end{equation}
Here   $r$ denotes the distance between coarse-grain
[P] particles,
$R_i = R_j = R_P=2.104$ \AA  \ is the effective
Born radius of phosphate particle.
The coefficient $C_1=14.400611$\AA eV,  $\epsilon_{out}=78$,
$\kappa=0.1$ what corresponds to physiological conditions. Parameter
$$
C_0=-\frac{1}{2} C_1\left(1- \frac{1}{\epsilon_{out}}\right) \sum_{i=0}^{N} \frac{1}{R_P}
$$
describes self-energy (solvation energy) of phosphate groups.

\begin{figure}[tb]
\includegraphics[angle=0, width=1\linewidth]{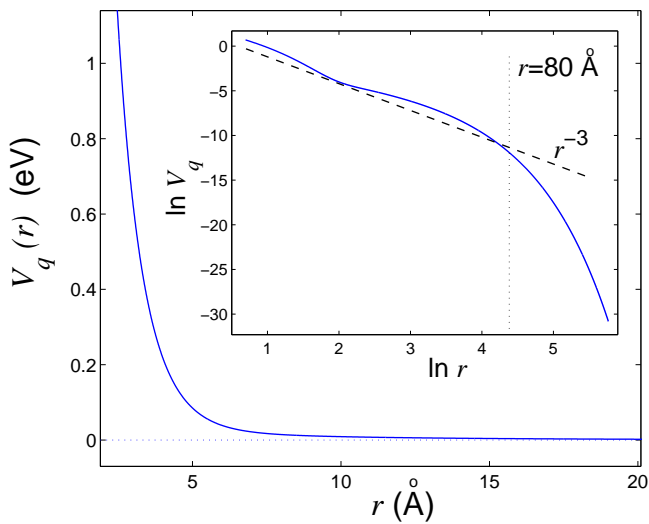}
\caption{\label{fig12}\protect
Electrostatics potential  $V_q(r)$, equation (\ref{f7}).
}
\end{figure}

The resulting total electrostatic potential due to a single
[P] particle as a function of distance
is shown in Fig.~\ref{fig12}. One can see that
for small distances
$r<80$\AA \ potential decreases with  increasing  distance $r$ as $r^{-3}$. For long
distances the fall-off  is exponential. Thus we can introduce a cut-off distance
$R_Q=100$\AA \ for the electrostatics interactions. For $r>R_Q$
interaction between particles is set to zero: $V_q=0$.

The last term $E_{vdW}$  in the Hamiltonian (\ref{f1})
describes van der Waals interaction between different side chain [P] and [C3] grains.
The potential depends on the distance $r$ between two grains and  is given by
$$
U_{ij}(r)=\epsilon_{ij}\left[\left(\frac{\sigma_{ij}}{r-d_{ij}}\right)^6-1\right]^2-\epsilon_{ij},
~i,j=P,C3,
$$
where
$\epsilon_{ij}=\sqrt{\epsilon_i\epsilon_j}$, $d_{ij}=d_i+d_j$, $\sigma_{ij}=\sigma_i+\sigma_j$,
energy parameters are $\epsilon_P=0.01$eV, $\epsilon_{C3}=0.005$eV,
diameters are $d_P=2.4$\AA, $d_{C3}=2$\AA, parameter
$\sigma_P=1.6$\AA, $\sigma_{C3}=1.9$\AA.

In practical applications of the 12CG model  one 
should keep in mind that the model was designed to describe 
only the double helical form of DNA, so it may not be appropriate 
to situation when melting or  base openings are expected. 
This limitation is the price one pays for computational efficiency: 
within our model van der Waals 
interactions  are calculated  only for backbone grains that 
belong to separate DNA strands, and only 
nearest neighbor base pairs interact. 

\bibliography{dna}
\end{document}